\documentclass[twocolumn,showpacs,preprintnumbers,amsmath,amssymb,floatfix]{revtex4}

\usepackage{graphicx}
\usepackage{dcolumn}
\usepackage{bm}

\newcommand{\beq}{\begin{equation}}
\newcommand{\eeq}{\end{equation}}
\newcommand{\bey}{\begin{eqnarray}}
\newcommand{\eey}{\end{eqnarray}}

\begin{document}

\title{Singularity-free dark energy star}

\author{Farook Rahaman}
\email{farook\_rahaman@yahoo.com} \affiliation{Department of
Mathematics, Jadavpur University, Kolkata 700 032, West Bengal,
India}
\author{Anil Kumar Yadav}
\email{abanilyadav@yahoo.co.in} \affiliation{Department of
Physics, Anand Engineering College, Keetham, Agra -282 007, India}
\author{Saibal Ray}
\email{saibal@iucaa.ernet.in} \affiliation{Department of Physics,
Government College of Engineering \& Ceramic Thechnology, Kolkata
700 010, West Bengal, India}
\author{Raju Maulick}
\email{rajuspinor@gmail.com} \affiliation{Department of
Mathematics, Jadavpur University, Kolkata 700 032, West Bengal,
India}
\author{Ranjan Sharma}
\email{rsharma@iucaa.ernet.in}\affiliation{Department of Physics,
P. D. Women's College, Jalpaiguri 735101, India.}

\date{\today}

\begin{abstract}
We propose a model for an anisotropic dark energy star where we
assume that the radial pressure exerted on the system due to the
presence of dark energy is proportional to the isotropic perfect
fluid matter density. We discuss various physical features of our
model and show
  that the model satisfies all the regularity conditions and is stable as well as singularity-free.
\end{abstract}


\maketitle

\section{Introduction}
Current cosmological observations of the accelerated expansion of
the universe strongly suggest that about 96\% of the total energy
content of the universe is exotic in nature out of which $73\%$ is
believed to be gravitationally repulsive in nature popularly
called {\it dark energy} and the remaining $23\%$ is attractive in
nature and exists in the form of dark matter
\cite{Perlmutter1998,Riess2004}. Consequently, cosmological models
based on dark energy either in the form of a cosmological constant
or in some other exotic forms of matter have got tremendous
attention in the recent past. From the astrophysical perspective,
if it is fundamentally impossible to get any observational
evidence for the existence of an event horizon in our universe
\cite{Abramowicz2002} (though our current understanding of the
general theory of relativity strongly favour the existence of
strong gravitational regions induced by compact objects covered
under the event horizon), one is tempted to look for alternative
models which may serve as alternatives to black holes. A dark
energy star is, in particular, interesting in this scenario
\cite{Lobo2006}.

Any interior solution to the vacuum Schwarzschild exterior
comprising a fluid distribution governed by an equation of
state(EOS) of the form $p = -\frac{1}{3} \rho$, may be considered
as a dark energy star \cite{Chapline2005}. In the past, various
model specific dark energy stars have been proposed (see for
example, Ref.
\cite{Chapline2005,Mazur2002,Lobo2006,Chan2008,Ghezzi2005} and
Ref. \cite{Paddy2008} for a recent review). In the present work,
we propose a model for an anisotropic dark energy star where we
assume that the radial pressure exerted on the system due to the
presence of dark energy is proportional to the isotropic perfect
fluid matter density. The stellar configuration comprises two
fluids - an ordinary baryonic perfect fluid together with an yet
unknown form of matter (dark energy) which is repulsive in nature.
We also assume that the two fluids are non interacting amongst
each other.

To describe the energy-momentum tensor for such a hybrid model, we
assume that our resulting composition is anisotropic in nature,
i.e. $p_r \neq p_t$, where $p_r$ and $p_t$ correspond to radial
and tangential pressure, respectively. Ever since the pioneering
works of Bowers and Liang \cite{Bowers1974}, anisotropic
relativistic stellar models have played an inportant role in the
description of compact stellar objects (see \cite{Herrera1995} for
a recent review). At the microscopic level, a variety of reasons
such as the existence of type $3A$ superfluid, phase transition,
pion condensation, rotation, magnetic field, mixture of two
fluids, bosonic composition etc., may give rise to anisotropic
pressures inside a stellar object. Recent observations on highly
compact astrophysical objects like X ray pulsar Her X-1, X ray
buster 4U 1820-30, millisecond pulsar SAX J 1808.4 - 3658, X ray
sources 4U 1728 - 34, etc., also strongly favour an anisotropic
matter distribution since the density inside such an ultra-compact
object is expected to be beyond nuclear matter density. In our
model, we assume that the ansisotropy is generated due to two
kinds of fluid distributions. In a repent paper
\cite{Yazadjiev2011}, an exterior solution corresponding to a two
fluid stellar model composed of non-interacting phantom scalar
field describing the dark energy and ordinary matter has been
reported which reduces to Schwarzschild solution in the absence of
the dark energy. In our paper, we match the interior solution to
the Schwarzschild exterior solution at the boundary.

Our paper is organized as follows: In Section II we have provided
the basic equations in connection to the proposed model for dark
energy star. Sections. III - VIII are dealt, respectively, with
the boundary conditions, TOV equation, energy conditions,
stability, mass-radius relation and junction conditions for the
solutions under consideration. Some concluding remarks are made in
the Section IX.

\section{Basic Equations and Their Solutions}
To describe the space-time of the dark energy stellar
configuration, we take the Krori and Barua \cite{Krori1975} metric
(henceforth KB) given by
\begin{equation}
ds^2=-e^{\nu(r)}dt^2 + e^{\lambda(r)}dr^2 +r^2
(d\theta^2 +sin^2\theta d\phi^2), \label{eq1}
\end{equation}
with $\lambda(r)=Ar^2$ and $\nu(r) = Br^2 + C$ where $A$, $B$ and
$C$ are arbitrary constants to be determined on  physical grounds.
The energy-momentum of the two fluids are such that
\begin{eqnarray}
T^{0}_{0} \equiv (\rho)_{eff} &=& \rho + \rho_{de}, \label{eq2}\\
T^{1}_{1} \equiv -(p_{r})_{eff} &=& -(p +p_{der}), \label{eq3}\\
T^{2}_{2} \equiv T^{3}_{3} \equiv -(p_{t})_{eff} &=& -(p+p_{det}), \label{eq4}
\end{eqnarray}
where $\rho$ and $p$ correspond to the energy density and pressure
of the baryonic matter, respectively, and $\rho_{de}$, $p_{der}$
and $p_{det}$ are the `dark' energy density, radial pressure and
tangential pressure, respectively. The left hand sides of
equations (\ref{eq2})-(\ref{eq4}) are the effective energy-density
and two pressures, respectively, of the composition.

The Einstein's field equations for the metric (\ref{eq1}) are then
obtained as (we assume $G = c=1$ under geometrized relativistic
units)
\begin{eqnarray}
\label{eq5}
8\pi\left(\rho+\rho_{de}\right) &=& e^{-\lambda}\left(\frac{\lambda^\prime}{r}-\frac{1}{r^2}\right) +
\frac{1}{r^2},\\
\label{eq6} 8\pi\left(p+p_{der}\right) &=&
e^{-\lambda}\left(\frac{\nu^\prime}{r}+\frac{1}{r^2}\right) -
\frac{1}{r^2},\\
\label{eq7}
8\pi\left(p+p_{det}\right) &=& \frac{e^{-\lambda}}{2}\left[\frac{{\nu^\prime}^2 - \lambda^{\prime}\nu^{\prime}}{2}
+\frac{\nu^\prime-\lambda^\prime}{r}+\nu^{\prime\prime}\right].
\end{eqnarray}

To solve the above set of equations, we assume that the dark
energy radial pressure is proportional to the dark energy density,
i.e.,
\begin{equation}
p_{der} = -\rho_{de}, \label{eq8}
\end{equation}
and the dark energy density is proportional to the matter density, i.e.,
\begin{equation}
\rho_{de} = \alpha\rho, \label{eq9}
\end{equation}
where $\alpha > 0$ is a proportionality constant. In connection to
the {\it ansatz} (\ref{eq8}) it is worthwhile to mention that the
equation of state of this type which implies that the matter
distribution under consideration is in tension is available in
literature and hence the matter is known as a `false vacuum' or
`degenerate vacuum' or
`$\rho$-vacuum'~\cite{Davies1984,Blome1984,Hogan1984,Kaiser1984}.

Now, from the metric (\ref{eq1}) we get $\lambda^\prime = 2Ar$,
$\nu^\prime = 2Br$ and $e^{-\lambda} = e^{-Ar^2}$ and substituting
these values in equations (\ref{eq5}) - (\ref{eq7}), together with
our assumptions as given in equations (\ref{eq8}) and (\ref{eq9}),
we get
\begin{eqnarray}
\label{eq10} 8\pi\rho &=&
\frac{1}{(1+\alpha)}\left[e^{-Ar^2}\left(2A-\frac{1}{r^2}\right)
+\frac{1}{r^2}\right],\\
\label{eq11} 8\pi(\rho +p) &=&
2 e^{-Ar^2}\left(A+B\right).
\end{eqnarray}

Subtracting equation (\ref{eq10}) from (\ref{eq11}), we get
\begin{multline}
\label{eq12} 8\pi p = e^{-Ar^2}\left(2A + 2B\right) -\\
\frac{1}{(1+\alpha)}\left[e^{-Ar^2}\left(2A-\frac{1}{r^2}\right)
+\frac{1}{r^2}\right].
\end{multline}

The equation (\ref{eq12}), alongwith (\ref{eq7}), then provides
\begin{multline}
\label{eq13}
 8\pi p_{det} = e^{-Ar^2}\left[B^{2}r^2  -ABr^2 -3A\right]\\
+
\frac{1}{(1+\alpha)}\left[e^{-Ar^2}\left(2A-\frac{1}{r^2}\right)
+\frac{1}{r^2}\right].
\end{multline}

Thus the effective energy density $(\rho)_{eff}$, effective radial
pressure $(p_{r})_{eff}$ and the effective tangential pressure
$(p_{t})_{eff}$ are obtained as
\begin{eqnarray}
(\rho)_{eff} &=&
\frac{1}{8\pi}\left[e^{-Ar^2}\left(2A-\frac{1}{r^2}\right)+\frac{1}{r^2}\right],\label{eq14}\\
(p_{r})_{eff} &=& \frac{1}{8\pi}\left[e^{-Ar^2}\left(2B+\frac{1}{r^2}\right)-\frac{1}{r^2}\right],\label{eq15}\\
(p_{t})_{eff} &=& \frac{1}{8\pi}\left[e^{-Ar^2}\left(B^2
r^2+2B-ABr^2-A\right)\right].\label{eq16}
\end{eqnarray}

Using equations (\ref{eq16}) - (\ref{eq18}) the
equation of state (EOS) corresponding to radial and
transverse directions may be written as
\begin{equation}
\label{eq17} \omega_r(r)   =
\frac{\left[e^{-Ar^2}\left(2B+\frac{1}{r^2}\right)-\frac{1}{r^2}\right]
}{\left[e^{-Ar^2}\left(2A-\frac{1}{r^2}\right)+\frac{1}{r^2}\right]}
\end{equation}
\begin{equation}
\label{eq18} \omega_t(r)   = \frac{\left[e^{-Ar^2}\left(B^2
r^2+2B-ABr^2-A\right)\right]
}{\left[e^{-Ar^2}\left(2A-\frac{1}{r^2}\right)+\frac{1}{r^2}\right]}
\end{equation}

It is interesting to note that the effective energy density
$(\rho_{eff})$, effective radial pressure $(p_{r~eff})$ and
effective tangential pressure $(p_{t~eff})$ are independent of
$\alpha$.

\begin{figure}[ptb]
\begin{center}
\vspace{0.2cm} \includegraphics[width=0.4\textwidth]{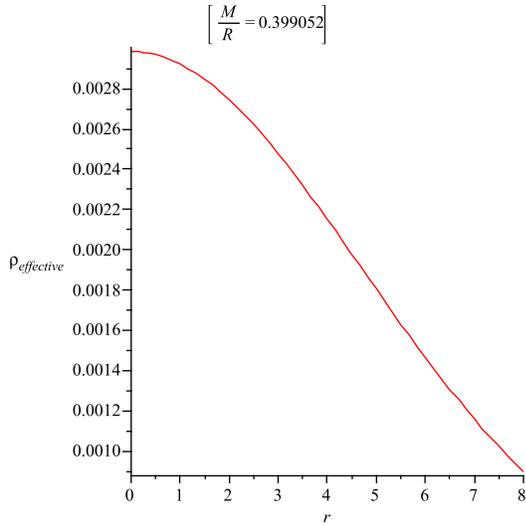}
\end{center}
\caption{The effective density parameter $\rho_{eff}$ is shown
against $r$.
}%
\label{fig1}
\end{figure}

\begin{figure}[ptb]
\begin{center}
\vspace{0.2cm} \includegraphics[width=0.4\textwidth]{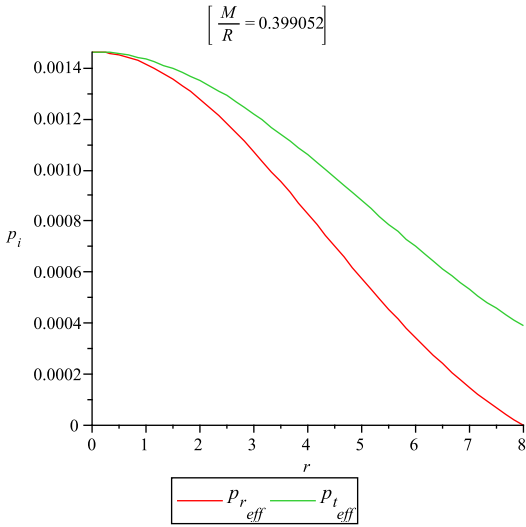}
\end{center}
\caption{Effective radial pressure and transverse pressures are
plotted against $r$.
}%
\label{fig2}
\end{figure}

We also note that
\[ \frac{d\rho_{eff}}{dr}= -\frac{1}{8\pi}\left[\left(4A^2 r
- \frac{2A}{r} -\frac{2}{r^3}\right)e^{-Ar^2}+\frac{2}{r^3}\right] < 0,\]\\
and
\[\frac{dp_{r~eff}}{dr} < 0.\]
We impose the following conditions for our anisotropic fluid
configuration to be physically acceptable:
\begin{itemize}
\item The density is positive definite and its gradient is negative everywhere within the fluid distribution.
\item The radial and tangential pressures are positive definite and the radial pressure gradient is negative definite.
\end{itemize}
The above results and Figs. \ref{fig1}-\ref{fig2} are in agreement with these
conditions.

\begin{figure}[ptb]
\begin{center}
\vspace{0.2cm} \includegraphics[width=0.4\textwidth]{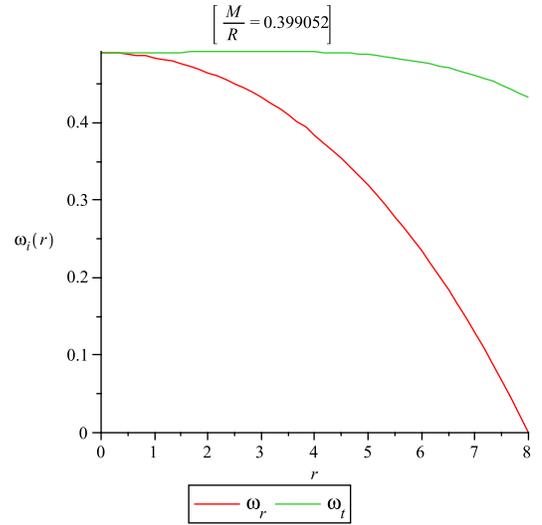}
\end{center}
\caption{The variation of the effective equation of state
parameter $\omega$ (radial and transverse) are shown against
$r$.
}%
\label{fig3}
\end{figure}

Note that, at $r=0$, our model provides\\
\[ \frac{d\rho_{eff}}{dr}=0,~~ \frac{dp_{r~eff}}{dr} = 0, \]
\[ \frac{d^2 \rho_{eff}}{dr^2}=-\frac{A^2}{\pi} < 0,\]
and
\[\frac{d^2 p_{r~eff}}{dr^2} < 0,\]
which indicate maximality of central density and central pressure.
Interestingly, similar to an ordinary matter distribution, the
bound on the effective EOS in this construction is given by $0<
\omega_i(r)<1$, (see Fig. \ref{fig3}) despite the fact that star
is constituted by the combination of ordinary matter and dark
energy.

\begin{figure}[ptb]
\begin{center}
\vspace{0.2cm} \includegraphics[width=0.4\textwidth]{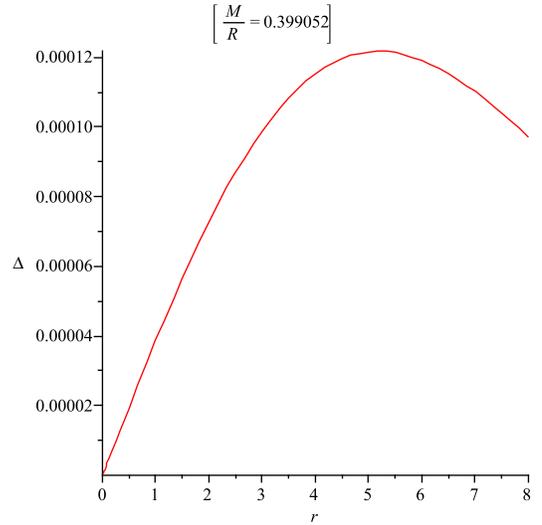}
\end{center}
\caption{The variation of the force, $\Delta  =
\frac{2}{r}\left(p_{t~eff}-p_{r~eff}\right)$ due to the local
anisotropy with respect to r.
}%
\label{fig4}
\end{figure}

The parameter $\Delta  = \frac{2}{r}\left(p_{t~eff}-p_{r~eff}\right)$ representing the
`force' due to the local anisotropy is obtained as
\begin{multline}
\label{eq19} \Delta  =
 \frac{1}{4\pi r
}\left[e^{-Ar^2}\left(B^2 r^2-ABr^2 -3A\right)\right]\\ +
\frac{1}{8\pi}\left[e^{-Ar^2}
\left(2A-\frac{1}{r^2}\right)+\frac{1}{r^2}\right].
\end{multline}
This `force' will be directed outward when $P_t > P_r$ i.e.
$\Delta > 0 $, and inward if $P_t < P_r$ i.e. $\Delta < 0$. As it
is apparent from the Fig.~\ref{fig4} of our model with a repulsive
`anisotropic' force ($\Delta > 0$) allows the construction of more
massive distributions.

\section{Boundary Conditions}
We match the interior metric to the Schwarzschild exterior
\begin{multline}
ds^{2}=-\left(1 - \frac{2M}{r}  \right)dt^2 + \left(1 -
\frac{2M}{r}    \right)^{-1}dr^2 \\+ r^2(d\theta^2+\sin^2\theta
d\phi^2), \label{eq20}
\end{multline}
at the boundary $r= R$. Assuming continuity of the metric functions $g_{tt}$, $g_{rr}$ and $\frac{\partial
g_{tt}}{\partial r}$ at the boundary surface $S$, we get
\begin{eqnarray}
1 - \frac{2M}{R}  &=& e^{BR^2+C},\label{eq21}\\
\left(1 - \frac{2M}{R}\right)^{-1}  &=& e^{AR^2}, \label{eq22}\\
\frac{M}{R^2} &=& BRe^{BR^2+C}.\label{eq23}
\end{eqnarray}

By solving Eqs.~(\ref{eq21})-(\ref{eq23}), we have
\begin{eqnarray}
A &=& - \frac{1}{R^2} \ln \left[ 1 - \frac{2M}{R} \right],\label{eq24}\\
B &=& \frac{1}{R^2} \left[\frac{M}{R}  \right] \left[1 -
\frac{2M}{R}   \right]^{-1}, \label{eq25}\\
C &=&  \ln \left[ 1 - \frac{2M}{R}  \right]- \frac{ \frac{M}{R}
 }{ \left[ 1 - \frac{2M}{R}
\right]}. \label{eq26}
\end{eqnarray}

We also impose the boundary conditions that at the boundary
$(p_r)_{eff}( r=R) = 0$ and $\rho_{~eff}( r=0) = b$ ($=a$
constant), where $b$ is the central density. Thus,
\begin{eqnarray}
A &=& \frac{8 \pi b}{3},\label{eq27} \\
B &=& \frac{1}{2R^2} \left[ e^{\frac{8 \pi b}{3} R^2} -1 \right].\label{eq28}
\end{eqnarray}

Combining, equations (\ref{eq24}) and (\ref{eq27}), we get
\begin{equation}
A = \frac{8 \pi b}{3} = - \frac{1}{R^2} \ln \left[ 1 - \frac{2M}{R} \right]. \label{eq29}
\end{equation}
Note that the values of $B$ obtained from equations
(\ref{eq25}) and (\ref{eq28}) are identical.

At this juncture, to get an insight of our model, let us first
evaluate some reasonable set for values of $A$, $B$, and $C$.
According to Buchdahl \cite{Buchdahl1959}, the maximum allowable
compactness (mass-radius ratio) for a fluid sphere is given by
$\frac{2M}{R} <\frac{8}{9}$. Accordingly, let us assume that we
have a dark energy star whose mass and radius are such that
$\frac{M}{R} = 0.3999052$. Due to highly compact nature of the
star, we set the radius of the star at $R = 8~$km. With these
specifications, we obtain the values of the constants $A$, $B$ and
$b$ as $A = .025$, $B = .030883$, $b = .002984$. Later, we have
shown that these values of $A$ and $B$ are justified since the
energy conditions imply  $ 2A \geq B \geq 0 $ (see Sec. V).

\section{TOV equation}
\begin{figure}[ptb]
\begin{center}
\vspace{0.2cm} \includegraphics[width=0.4\textwidth]{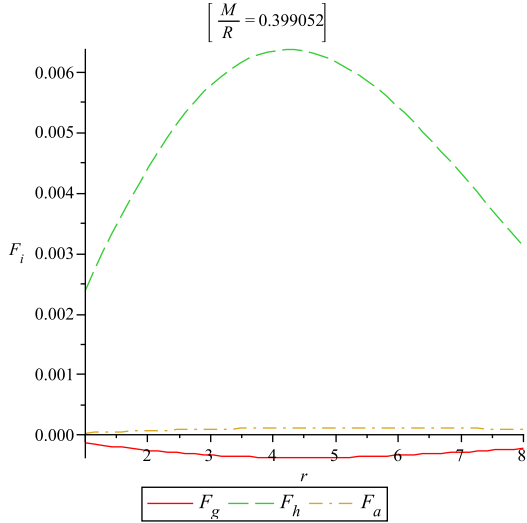}
\end{center}
\caption{Three different forces acting on fluid elements in static
equilibrium is shown against $r$.
}%
\label{fig5}
\end{figure}

For an anisotropic fluid distribution, the generalized TOV equation is given by
\begin{equation}
\label{eq30} \frac{d(p_{r ~eff})}{dr}   +
\nu^\prime\left(\rho_{eff} +p_{r~eff}\right) +
\frac{2}{r}\left(p_{r~eff} - p_{t~eff}\right) = 0.
\end{equation}
Following Ponce de Le\'{o}n \cite{Leon1993}, we write
the above TOV equation as
\begin{multline}
-\frac{M_G\left(\rho_{eff}+p_{r~eff}\right)}{r^2}e^{\frac{\lambda-\nu}{2}}-\frac{dp_{r~eff}}{dr}\\
+\frac{2}{r}\left(p_{t~eff}-p_{r~eff}\right)=0, \label{eq31}
\end{multline}
where $M_G=M_G(r)$ is the effective gravitational mass inside a
sphere of radius $r$ and is given by
\begin{equation}
M_G(r)=\frac{1}{2}r^2e^{\frac{\nu-\lambda}{2}}\nu^{\prime},\label{eq32}
\end{equation}
which can easily be derived from the Tolman-Whittaker formula and
the Einstein's field equations. Obviously, the modified TOV
equation describes the equilibrium condition for the dark star
subject to gravitational and hydrostatic plus another force due to
the anisotropic nature of the stellar object. Using equations
(\ref{eq14}) - (\ref{eq16}), the above equation can be written as
\begin{equation}
 F_g+ F_h + F_a=0,\label{eq33}
\end{equation}
where,
\begin{eqnarray}
F_g &=& -B r\left(\rho_{eff}+p_{r~eff}\right),\label{eq34}\\
F_h &=& -\frac{dp_{r~eff}}{dr},\label{eq35}\\
F_a &=& \frac{2}{r}\left(p_{t~eff} -p_{r~eff}\right).\label{eq36}
\end{eqnarray}
The profiles of $F_g$, $F_h$ and $F_a$ for our chosen source are
shown in Fig.~\ref{fig5}. The figure indicates that the static
equilibrium is attainable due to pressure anisotropy,
gravitational and hydrostatic  forces.

\section{Energy Conditions}
\begin{figure}[ptb]
\begin{center}
\vspace{0.2cm}
\includegraphics[width=0.4\textwidth]{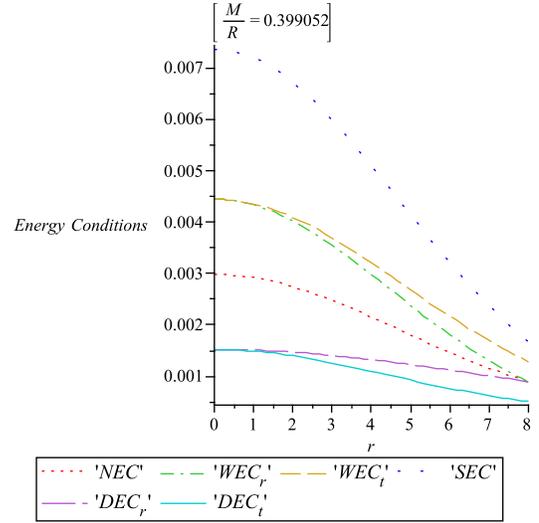}
\end{center}
\caption{The variation of left hand side of the expressions of
energy conditions are shown against $r$.
}%
\label{fig6}
\end{figure}

In this section, we verify whether our particular choices of the
values of mass and radius leading to solutions for the unknownn
parameters, satisfy the following conditions through out the
configuration:
\[\rho_{eff}  \geq 0, \]
\[\rho_{eff}+p_{r~eff}  \geq 0, \]
\[\rho_{eff}+p_{t~eff}  \geq 0, \]
\[\rho_{eff}+p_{r~eff}+2p_{t~eff}  \geq 0, \]
\[\rho_{eff} > |p_{r~eff}|,\]
\[\rho_{eff} > |p_{t~eff}|.\]
Note that all the energy conditions namely, the null energy
condition (NEC), weak energy condition (WEC), strong energy
condition (SEC) and dominant energy condition (DEC), for our
particular choices of the values of mass and radius, are satisfied
as shown in Fig.~\ref{fig6}. It is interesting to note here that
the model satisfies the strong energy condition, which implies
that the space-time does contain a black hole region.

The anisotropy, as expected, vanishes at the centre i.e.,
$p_{t~eff} = p_{r~eff} = p_{0~eff}=\frac{2B-A}{8\pi}$ at r=0. The
effective energy density and the two pressures are also well
behaved in the interior of the stellar configuration.

Employing the energy conditions at the centre ($r=0$), we may get
a bound on the constants $A$ and $B$ as follows:
\\
(i) NEC: $p_{0~eff}+\rho_{0~eff}\geq0$ $\Rightarrow$ $A+B\geq0$,\\
\\
(ii) WEC: $p_{0~eff}+\rho_{0~eff}\geq0$ $\Rightarrow$ $A+B\geq0$, \\
\\
$\rho_{0~eff}\geq0$ $\Rightarrow$ $A\geq0$,\\
\\
(iii) SEC: $p_{0~eff}+\rho_{0~eff}\geq0$ $\Rightarrow$ $A+B\geq0$,\\
\\
$3p_{0~eff}+\rho_{0~eff}\geq0$ $\Rightarrow$ $B\geq0$,\\
\\
(iv) DEC: $\rho_{0~eff} > |p_{0~eff}|$ $\Rightarrow$ $2A\geq B$.\\

\section{Stability}
\begin{figure}[ptb]
\begin{center}
\vspace{0.2cm} \includegraphics[width=0.4\textwidth]{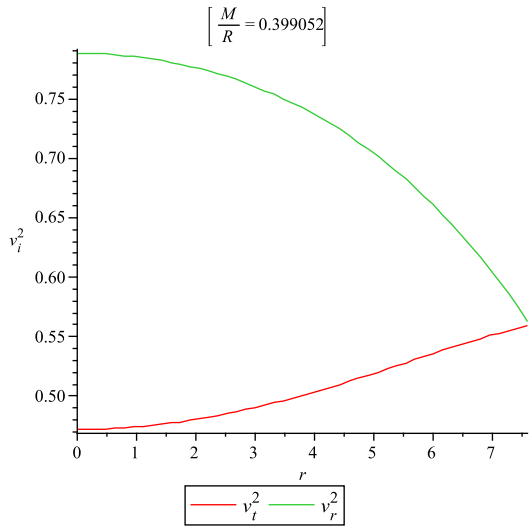}
\end{center}
\caption{The variation of radial sound speed $v_{sr}^2 $ and
tangential sound speed $v_{st }^2 $ are shown against $r$.
}%
\label{fig7}
\end{figure}

For a physically acceptable model, one expects that the velocity of sound
should be within the range $0 \leq v_s=(\frac{dp}{d\rho}) \leq 1$ \cite{Herrera1992,Abreu2007}.
In our anisotropic model, we define sound speeds as
\begin{multline}
\label{eq37}
v_{sr}^{2}=\frac{dp_{r~eff}}{d\rho_{eff}}\\=-1+\frac{4Are^{-Ar^2}(A+B)}{\left(2A-\frac{1}{r^2}\right)2Are^{-Ar^2}+\frac{2}{r^3}\left(1-e^{-Ar^2}\right)},
\end{multline}

\begin{multline}
\label{eq38}
v^2_{st}=\frac{dp_{t~eff}}{d\rho_{eff}}\\=\frac{e^{-Ar^2}\left[2Ar\left(B^2
r^2+2B-ABr^2-A\right)+2Br(A-B)\right]}{\left(2A-\frac{1}{r^2}\right)2Are^{-Ar^2}+\frac{2}{r^3}
\left(1-e^{-Ar^2}\right)}.
\end{multline}
We plot the radial and transverse sound speeds in Fig.~\ref{fig7}
and observe that these parameters satisfy the inequalities $0\leq
v_{sr}^2 \leq 1$ and $0\leq v_{st}^2 \leq 1$ everywhere within the
stellar object.

Equations (\ref{eq37}) and (\ref{eq38}) lead to
\begin{multline}
\label{eq39} v^2_{st}-v^2_{sr}\\=1-\frac{e^{-Ar^2}\left[2A^2 B r^3
+6A^{2}r + 2Br^2 -2ABr - 2AB^2
r^3\right]}{\left(2A-\frac{1}{r^2}\right)2Are^{-Ar^2}+\frac{2}{r^3}
\left(1-e^{-Ar^2}\right)}.
\end{multline}
From equation (\ref{eq39}), we note that $v^2_{st}-v^2_{sr}\leq 1$
. Since, $0\leq v_{sr}^2 \leq 1$ and $0\leq v_{st}^2 \leq 1$,
therefore,  $\mid v_{st}^2 - v_{sr}^2 \mid \leq 1 $. In
Fig.~\ref{eq8}, we have plotted  $\mid v_{st}^2 - v_{sr}^2 \mid$.

Now, to examine the stability of local anisotropic matter
distribution, we use Herrera's \cite{Herrera1992} cracking (or
overturning) concept which states that the region for which radial
speed of sound is greater than the transverse speed of sound is a
potentially stable region. Thus, if the difference of the two
sound speeds $v_{st}^2 - v_{sr}^2 $ retains the same sign
everywhere within a matter distribution, no cracking will occur.
In our case, Fig.~\ref{fig9} indicates that there is no change of
sign for the term $v_{st}^2 - v_{sr}^2 $ within the specific
configuration since the difference is negative throughout the
distribution. Therefore, we conclude that our dark energy star
model is stable.

\begin{figure}[ptb]
\begin{center}
\vspace{0.2cm}
\includegraphics[width=0.4\textwidth]{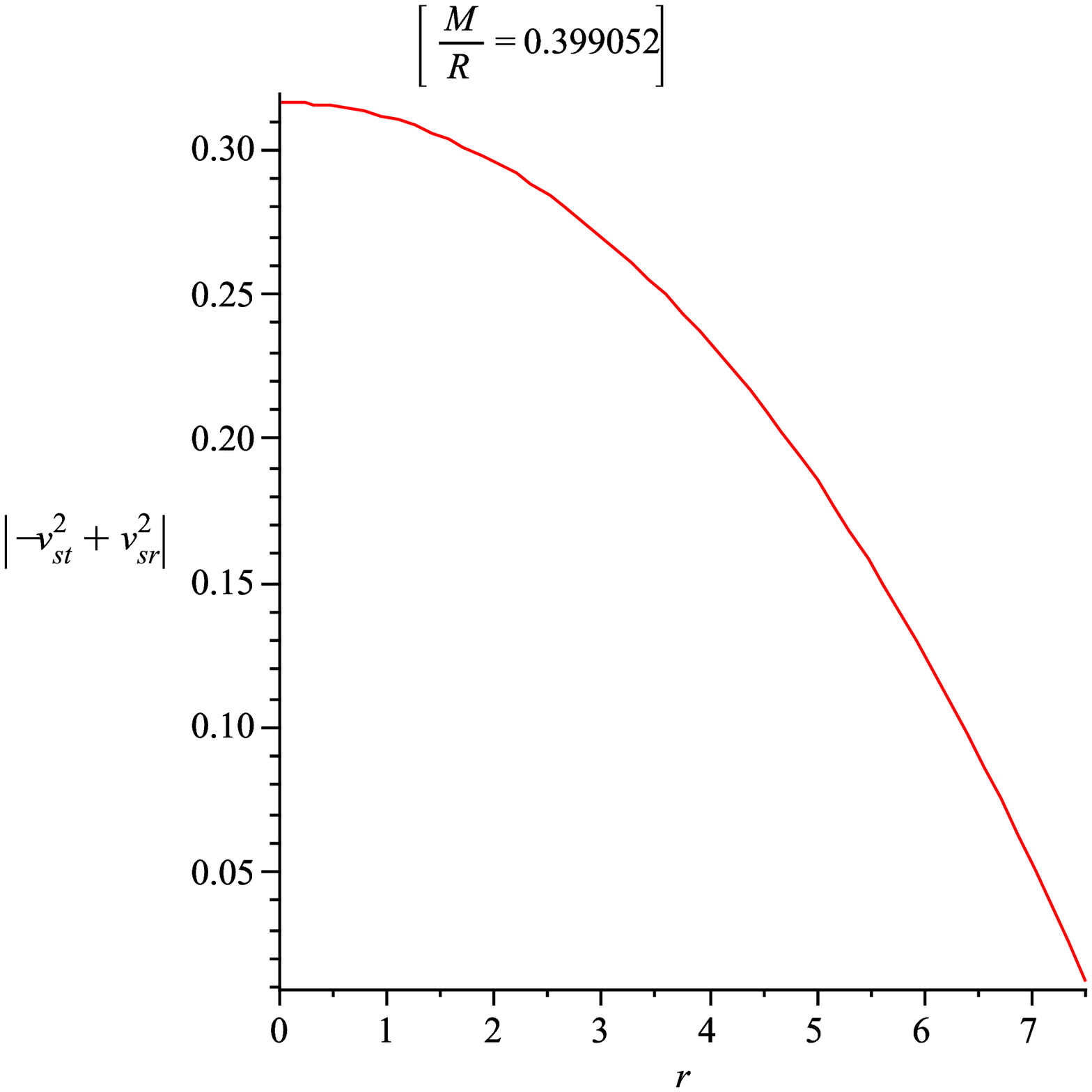}
\end{center}
\caption{The variation of $\mid v_{st}^2 - v_{sr}^2 \mid $ is
shown against $r$.
}%
\label{fig8}
\end{figure}

\begin{figure}[ptb]
\begin{center}
\vspace{0.2cm} \includegraphics[width=0.4\textwidth]{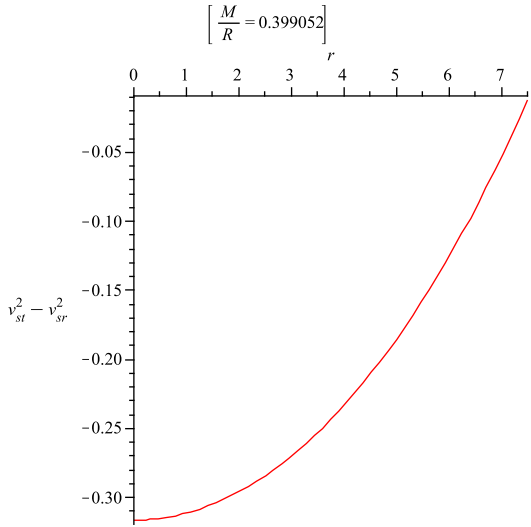}
\end{center}
\caption{The variation of $ v_{st}^2 - v_{sr}^2 $ is shown against
$r$.
}%
\label{fig9}
\end{figure}

\section{Mass-Radius relation}
\begin{figure}[ptb]
\begin{center}
\vspace{0.2cm} \includegraphics[width=0.4\textwidth]{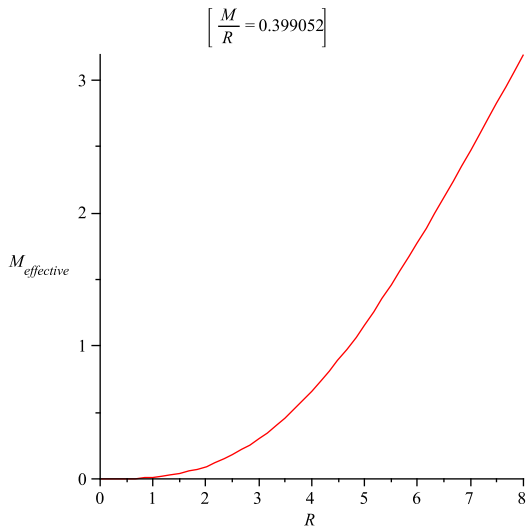}
\end{center}
\caption{The variation of $M_{effective}$  is shown against R.
}%
\label{fig10}
\end{figure}

\begin{figure}[ptb]
\begin{center}
\vspace{0.2cm}
\includegraphics[width=0.4\textwidth]{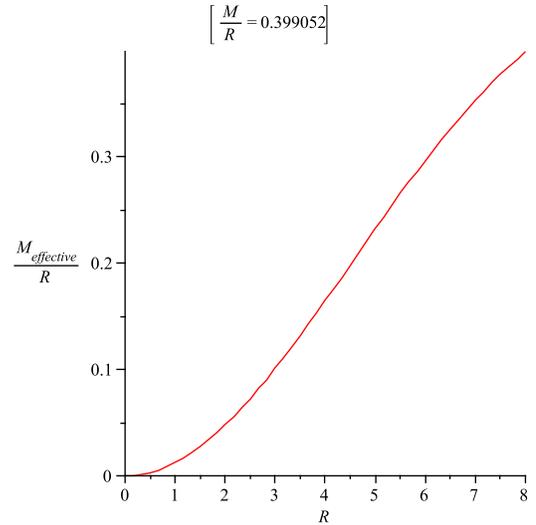}
\end{center}
\caption{The variation of $\frac{M _{effective}}{R}$  is shown
against R.
}%
\label{fig11}
\end{figure}

In this section, we study the maximum allowable mass-radius ratio
in our model. For a static spherically symmetric perfect fluid
star, Buchdahl \cite{Buchdahl1959} showed that the maximally
allowable mass-radius ratio is given by $\frac{2M}{R} <
\frac{8}{9}$ (for a more generalized expression for the same see
Ref. \cite{Mak2001}). In our model, the effective gravitational
mass in terms of the effective energy density $\rho_{eff}$ can be
expressed as
\\
\begin{multline}
\label{eq40}
M_{eff}=4\pi\int^{R}_{0}\left(\rho+\rho_{de}\right)r^2 dr =
\frac{1}{2}R\left( 1-e^{-AR^2}
 \right).
\end{multline}
In Fig.~\ref{fig10}, we plot this mass-radius relation. We have
also plotted $\frac{M _{eff}}{R}$ against $R$ (see
Fig.~\ref{fig11}) which shows that the ratio $\frac{M _{eff}}{R}$
is an increasing function of the radial parameter. We note that a
constraint on the maximum allowed mass-radius ratio in our case is
similar to the isotropic fluid sphere, i.e., $\frac{M}{R} <
\frac{4}{9}$ as obtained earlier. The compactness of the star is
given by
\begin{equation}
\label{eq41} u= \frac{ M_{eff}(R)} {R}=  \frac{1}{2}\left(
1-e^{-AR^2}
 \right).
\end{equation}

The surface redshift ($Z_s$) corresponding to
the above compactness ($u$) is obtained as
\begin{equation}
\label{eq42} Z_s= ( 1-2 u)^{-\frac{1}{2}} - 1,
\end{equation}
where
\begin{equation}
\label{eq43} Z_s=  e^{ \frac{A}{2}R^2}.
\end{equation}
Thus, the maximum surface redshift for an anisotropic star of
radius $8~$km turns out to be $Z_s = 2.225541$.

\begin{figure}[ptb]
\begin{center}
\vspace{0.2cm}
\includegraphics[width=0.4\textwidth]{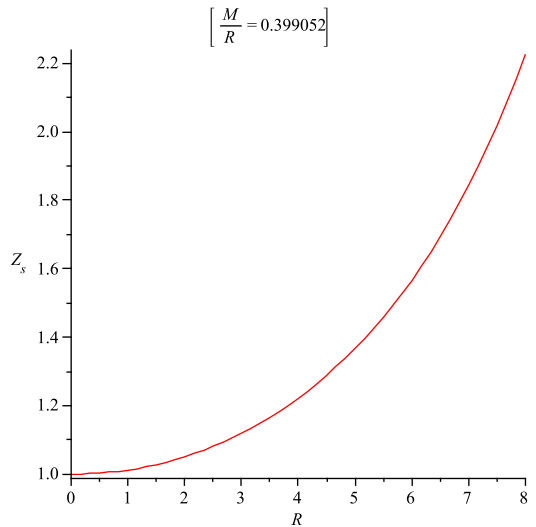}
\end{center}
\caption{The variation of redshift function $Z_s$  is shown
against R.
}%
\label{fig12}
\end{figure}

\section{Junction Condition}
One of the issues in connection with a static anisotropic matter
distribution is that, though the radial pressure at the boundary
of the star must vanish, the tangential pressure is not
necessarily zero at the boundary. This forces us to examine the
junction conditions of a static anisotropic star in closer
details. We propose here a  shell type envelope at the boundary
surface so as to address this issue. Note that the fundamental
junction condition for a static star is that there has to be a
smooth matching between the interior solution and Schwarzschild
exterior at the boundary. Now, though the metric coefficients must
be continuous at the junction surface $S$ where $r=R$, their
derivatives may not be continuous at the junction. In other words,
the affine connections may be discontinuous at the boundary
surface. This can be taken care of if we consider the second
fundamental forms of the boundary shell. The second fundamental
forms associated with the two sides of the shell
\cite{Israel1966,Usmani2010,Rahaman2010b,Rahaman2011} are given by
\begin{equation}K_{ij}^\pm = - n_\nu^\pm\ \left[\frac{\partial^2X_\nu}
{\partial \xi^i\partial \xi^j } +
 \Gamma_{\alpha\beta}^\nu \frac{\partial X^\alpha}{\partial \xi^i}
 \frac{\partial X^\beta}{\partial \xi^j }\right]_{ |_S}, \label{eq44}
 \end{equation}
where $ n_\nu^\pm\ $ are the unit normals to $S$ and can be
writtten as
\begin{equation} n_\nu^\pm =  \pm   \left| g^{\alpha\beta}\frac{\partial f}{\partial X^\alpha}
 \frac{\partial f}{\partial X^\beta} \right|^{-\frac{1}{2}} \frac{\partial f}{\partial X^\nu},\label{eq45}
 \end{equation}
with $ n^\mu n_\mu = 1$. In Eq.~(\ref{eq45}), $\xi^i$ are the
intrinsic coordinates on the shell with $f =0$ being the
parametric equation of the shell $S$ and $-$ and $+$ correspond to
interior and exterior (Schwarzschild) metrices. Note that radial
pressure on the shell is zero. By using Lanczos equations
\cite{Israel1966,Usmani2010,Rahaman2010b,Rahaman2011}, the surface
energy term $\Sigma$ and surface tangential pressures $ p_\theta =
p_\phi \equiv p_t$ may be obtained as
\begin{eqnarray}
\Sigma &=&  - \frac{1}{4\pi R}\left[ \sqrt{e^{-\lambda}}\right]_-^+, \label{eq46}\\
p_t &=&   \frac{1}{8\pi R}\left[ \left( 1 + \frac{R \nu^\prime
}{2}\right) \sqrt{e^{-\lambda}}\right]_-^+.\label{eq47}
\end{eqnarray}
Since the metric functions are continuous on $S$, we have
\begin{equation}
\Sigma = 0,\label{eq48}
\end{equation}
and
\begin{multline}
p_t = \frac{1}{8\pi R} \left[ \frac{\left(\frac{M}{ R}  \right)}{
{\sqrt{1 - \frac{2M}{R} }}} -BR^2\sqrt{e^{-AR^2}} \right].\label{eq49}
\end{multline}
Therefore, we are now in a position to match our interior solution
to the Schwarzschild exterior in the presence of a thin shell.

\begin{figure}[ptb]
\begin{center}
\vspace{0.2cm} \includegraphics[width=0.4\textwidth]{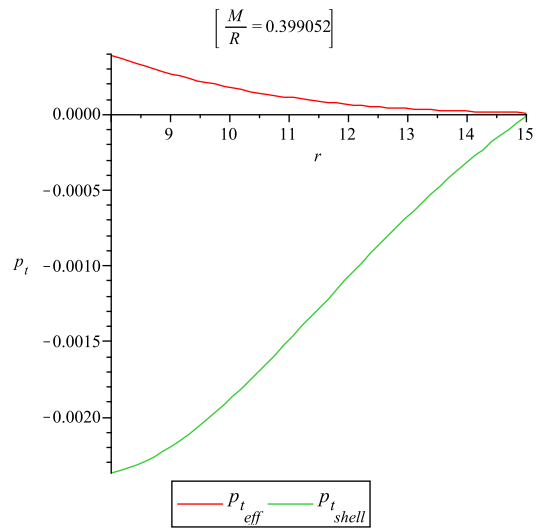}
\end{center}
\caption{Plot for $p_{t~ eff}$ and $p_{t ~shell} $ is shown
against r.
}%
\label{fig13}
\end{figure}

From Fig.~\ref{fig13}, we note that the effective transverse
pressure at the boundary ($R=8~$km) is positive though the
transverse pressure in the shell is negative. This clearly
indicates that the static equilibrium may be attained due to
positive $p_{t~ eff}$ and negative $p_{t ~shell}$. In this figure,
we have plotted $p_{t~ eff}$ and $p_{t ~shell}$ together by
assuming that the width of the shell is in between $8-15~$km. We
observe that at $r=15~$km, both $p_{t~ eff}$ and $p_{t ~shell}$
vanish simultaneously. Thus the thickness of the shell in this
case is $7~$km. Physically, this suggests that anisotropic matter
is confined within $8~$km from the centre of the star and the
outer region contains a thick shell extending up to $7~$km. The
thick shell is characterized by zero energy density and non zero
transverse pressure though the shell does not exert any radial
pressure.

\section{Conclusion}
Dark energy stellar models have found astrophysical relevance for
various reasons, one particular reason being it's importance as an
alternative candidate to a black hole. The model developed here
satisfies all the physical requirements and is horizon free and,
therefore, can potentially describe a compact object which is
neither a neutron stars nor a quark star.

Similar to many model proposed earlier, our model also requires an
envelope just outside the baryonic matter for smooth matching with
the Schwarzschild exterior space-time. For matching we have {\it
firstly}, assumed  continuity of the metric functions $g_{tt}$,
$g_{rr}$ and $\frac{\partial g_{tt}}{\partial r}$ at the boundary
surface $S$, and {\it secondly} imposed the boundary conditions
that at the boundary $(p_r)_{eff}( r=R) = 0$ and $\rho_{~eff}(
r=0) = b$ ($=a$ constant), where $b$ is the central density. Thus
we get a set of expressions for $A$, $B$ and $C$ which, due to the
maximum allowable compactness for a fluid sphere
\cite{Buchdahl1959}, eventually can be worked out as $A = .025$,
$B = .030883$, $b = .002984$ for the assumed mass-radius ratio as
$\frac{M}{R} = 0.3999052$. Later on, in Sec. V, it has been shown
that these values of $A$ and $B$ are justified since the energy
conditions imply $ 2A \geq B \geq 0 $.

Regarding stability of local anisotropic matter distribution, we
use cracking concept of Herrera \cite{Herrera1992} which states
that the region for which radial speed of sound is greater than
the transverse speed of sound is a potentially stable region. It
is observed from Fig.~\ref{fig9} that there is no change of sign
for the term $v_{st}^2 - v_{sr}^2 $ within the specific
configuration and hence advocating in favour of stability of our
dark energy star model.

Let us now concentrate on some of the other works on KB analysis,
especially the works by Varela et al. \cite{Varela2010}  and
Farook et al. \cite{Farook2010a}. In both the works static,
spherically symmetric, Einstein-Maxwell spacetime have been
considered with a fluid source of anisotropic stresses whereas the
present investigation is neutral one with anisotropic fluid
source. However, a common feature of all these KB-models is
singularity-free, stable configurations. Valera et
al.\cite{Varela2010} in their work have found out a link of their
construction with a charged strange quark star as well as models
of dark matter including massive charged particles. Farook et
al.\cite{Farook2010a}, by using a Chaplygin-type EOS, predicted
the possible existence of a Chaplygin charged dark energy star or
a strange quark star of radius about 8 km.

Therefore, it is interesting to note that present work deals with
a singularity-free spherically symmetric body of radius $r=15$~km
such that both $p_{t~ eff}$ and $p_{t ~shell}$ vanish
simultaneously. The model physically contains anisotropic matter
which is confined within $8~$km from the centre of the star and
the outer region contains a thick shell extending up to $7~$km.
Here the thick shell is characterized by zero energy density and
non-zero transverse pressure though the shell does not exert any
radial pressure. Note that in a recently proposed toy model
\cite{Dzhunushaliev2011}, a relativistic stellar configuration has
been developed where  the core of the star is characterized by a
wormhole like solution for some kind of exotic matter violating
the weak/null energy condition and is surrounded by some ordinary
matter satisfying a polytropic EOS. This kind of theoretical
modelling would get observational support in the future.

We hope our model inspires observational workers to search this
type of stars. That is, we mean,  the stars containing anisotropic
matter which is confined within certain radius from the centre of
the star and the outer region contains a thick shell extending up
to several kilometer.

\section*{Acknowledgments}
FR, SR and RS  gratefully acknowledge support from IUCAA, Pune,
India under Visiting Associateship under which a part of this work
was carried out. FR is also thankful to PURSE for providing
financial support. We thank the anonymous referee for drawing our
attention to a couple of references relevant to our studies.


\begin{thebibliography}{99}

\bibitem[\protect\citeauthoryear{Perlmutter}{1998}]{Perlmutter1998} S. Perlmutter {\em et al}, Nature {\bf 391}, 51 (1998).

\bibitem[\protect\citeauthoryear{Riess}{2004}]{Riess2004} A. G. Riess {\em et al}, Astrophys. J. {\bf 607}, 665 (2004).

\bibitem[\protect\citeauthoryear{Abramowicz}{2002}]{Abramowicz2002} M. A. Abramowicz, W. Kluzniak and J. P. Lasota, Astron. Astrophys. {\bf 396}, L31 (2002).

\bibitem[\protect\citeauthoryear{Lobo}{2006}]{Lobo2006} F. S. N. Lobo, Class. Quantum Grav. {\bf 23}, 1525 (2006), {\bf24}, 1069 (2007).

\bibitem[\protect\citeauthoryear{Chapline}{2005}]{Chapline2005} G. Chapline, Proceedings of the Texas Conference on Relativistic Astrophysics, Stanford, CA, December, (2004).

\bibitem[\protect\citeauthoryear{Mazur}{2002}]{Mazur2002} P. O. Mazur and E. Mottola,
arXiv:gr-qc/0109035.

\bibitem[\protect\citeauthoryear{Chan}{2008}]{Chan2008} R. Chan, M. F. A. da Silva and J. F. V. da Rocha,
Gen. Relativ. Gravit. {\bf 41}, 1835 (2009).

\bibitem[\protect\citeauthoryear{Ghezzi}{2005}]{Ghezzi2005} C. R. Ghezzi, Phys. Rev. D {\bf 72} 104017 (2005).

\bibitem[\protect\citeauthoryear{Padmanabhan}{2008}]{Paddy2008} T. Padmanabhan, Gen. Relativ. Gravit. {\bf40}, 529 (2008).

\bibitem[\protect\citeauthoryear{Bowers}{1974}]{Bowers1974} R. L. Bowers and E. P. T. Liang, Astrophys. J., {\bf188}, 657 (1974).

\bibitem[\protect\citeauthoryear{Herrera}{1995}]{Herrera1995} L. Herrera and N. O. Santos,  Astrophys. J., {\bf438}, 308 (1995).
\bibitem[\protect\citeauthoryear{Yazadjiev}{2011}]{Yazadjiev2011} Stoytcho S. Yazadjie, arXiv:1104.1865v2 [gr-qc].


\bibitem[\protect\citeauthoryear{Krori}{1975}]{Krori1975} K. D. Krori and J. Barua, J. Phys. A.: Math. Gen. {\bf 8},
508 (1975).

\bibitem[\protect\citeauthoryear{Davies}{1984}]{Davies1984} C. W. Davies, Phys. Rev. {\bf D30}, 737 (1984).

\bibitem[\protect\citeauthoryear{Blome}{1984}]{Blome1984} J. J. Blome and W. Priester, Naturwissenshaften {\bf 71}, 528 (1984).

\bibitem[\protect\citeauthoryear{Hogan}{1984}]{Hogan1984} C. Hogan, Nature {\bf 310}, 365 (1984).

\bibitem[\protect\citeauthoryear{Kaiser}{1984}]{Kaiser1984} N. Kaiser and A. Stebbins, Nature {\bf 310}, 391 (1984).

\bibitem[\protect\citeauthoryear{Buchdahl}{1959}]{Buchdahl1959} H. A. Buchdahl, Phys. Rev. {\bf 116}, 1027 (1959).

\bibitem[\protect\citeauthoryear{leon}{1993}]{Leon1993} J. Ponce de Le\'{o}n, Gen. Relativ. Gravit.
{\bf 25}, 1123 (1993).

\bibitem[\protect\citeauthoryear{Herrera}{1992}]{Herrera1992} L. Herrera, Phys. Lett. A, {\bf 165} 206, (1992).

\bibitem[\protect\citeauthoryear{Abreu}{2007}]{Abreu2007} H. Abreu, H. Hernandez and L. A. Nunez, Class. Quantum Gravit. {\bf 24}, 4631 (2007).

\bibitem[\protect\citeauthoryear{Mak}{2001}]{Mak2001} M. K. Mak, P. N. Dobson and T. Harko, Europhys. Lett. {\bf 55}, 310 (2001).

\bibitem[\protect\citeauthoryear{Israel}{1966}]{Israel1966} W. Israel, Nuo. Cim. B {\bf 44}, 1 (1966);
erratum - ibid. {\bf 48B}, 463 (1967).

\bibitem[\protect\citeauthoryear{Usmani}{2010}]{Usmani2010} A. A. Usmani, Z. Hasan, F. Rahaman, Sk. A.
Rakib, S. Ray, P. K. F. Kuhfittig, Gen. Relativ. Gravit. {\bf 42}, 2901 (2010).

\bibitem[\protect\citeauthoryear{Rahaman}{2010}]{Rahaman2010b} F. Rahaman, K. A. Rahman, Sk. A Rakib, P. K. F.
Kuhfittig, Int. J. Theor. Phys. {\bf 49}, 2364 (2010).

\bibitem[\protect\citeauthoryear{Rahaman}{2011}]{Rahaman2011} F. Rahaman, P. K. F. Kuhfittig, M. Kalam, A. A. Usmani and S. Ray, arXiv:gr-qc/1011.3600.

\bibitem[\protect\citeauthoryear{Varela}{2010}]{Varela2010} V. Varela, F. Rahaman, S. Ray, K. Chakraborty and M. Kalam, Phys. Rev. D {\bf 82}, 044052 (2010).

\bibitem[\protect\citeauthoryear{Farook}{2010}]{Farook2010a} F. Rahaman, S. Ray, A. K. Jafry and K. Chakraborty, Phys. Rev. D {\bf 82}, 104055 (2010).
\bibitem[\protect\citeauthoryear{Dzhunushaliev}{2011}]{Dzhunushaliev2011} Vladimir Dzhunushaliev et al,  arXiv:1102.4454v3 [astro-ph.GA]


\end{thebibliography}
\end{document}